\documentclass[a4paper]{article}

\usepackage{17lomcon}        
\usepackage{cite}            
\usepackage{epsfig}          
\usepackage{amssymb}

\newcommand{\eV}{\ensuremath{\, \mathrm{eV}}}
\newcommand{\lsn}{LS$\nu$}
\newcommand{\Neff}{\ensuremath{N_{\mathrm{eff}}}}
\newcommand{\DNeff}{\ensuremath{\Delta\Neff}}
\newcommand{\meff}[1]{\ensuremath{m^{\mathrm{eff}}_{#1}}}
\newcommand{\lcdm}{$\Lambda$CDM}

\begin{document}


\title{LIGHT STERILE NEUTRINOS IN COSMOLOGY}

\author{Stefano Gariazzo \email{gariazzo@to.infn.it}}

\affiliation{Department of Physics, University of Torino,
and INFN, Sezione di Torino, Via P. Giuria 1, I--10125 Torino, Italy}


\date{}
\maketitle


\begin{abstract}
We briefly discuss
the main effects of the presence of a Light Sterile Neutrino
(\lsn) in Cosmology
and how its properties can be constrained by Cosmic Microwave Background (CMB)
and other cosmological measurements.
\end{abstract}

\section{Introduction}
The measurements of neutrino oscillations in
atmospheric, solar and long-baseline
neutrino oscillation experiments
firmly indicate that at least two neutrino mass eigenstates
have a positive mass
(see e.g.\ Refs.~\cite{Giunti:2007ry,Bilenky:2010zza,PDG-2014}).
The mixing between the flavour neutrino eigenstates,
that we indicate with $\nu_\alpha$ ($\alpha=e,\mu,\tau$),
and the mass neutrino eigenstates,
indicated with $\nu_k$ ($k=1,2,3$),
can be written in terms of the mixing matrix $U_{\alpha k}$.
The mixing relation is
$\nu_\alpha=\sum U_{\alpha k} \nu_k$.
The mixing matrix $U_{\alpha k}$ is described by three mixing angles,
a CP violating phase and two Majorana phases,
that are physical only if neutrino are Majorana particles.
The description of the three neutrino mixing is completed
by the existence of two squared-mass differences
$\Delta m_{SOL}^2$ and $\Delta m_{ATM}^2$, which are measured
in neutrino oscillation experiments
(see e.g.\ Ref.~\cite{PDG-2014}).

Among the numerous neutrino oscillation experiments,
it is possible to find
some anomalies that cannot be explained in the context of the
three neutrino mixing.
These anomalies are related to the results obtained
in short-baseline (SBL) neutrino oscillation experiments
by the LSND experiment \cite{Aguilar:2001ty},
by several experiments measuring
the electron antineutrino flux from nuclear reactors
\cite{Mention:2011rk}
and in the calibration of the GALLEX and SAGE Gallium
solar neutrino experiments \cite{Giunti:2006bj}
(see Ref.~\cite{Gariazzo:2015rra}
for a full list of references).
A possible explanation of these anomalies involves new neutrino
oscillations, driven by a third squared-mass difference
$\Delta m^2_{SBL}\simeq 1\eV$,
that implies the existence of a fourth
neutrino mass eigenstate.
This additional \emph{light}
neutrino degree of freedom should be mainly
\emph{sterile}.
We assume
that the active neutrinos have a negligible mass
with respect to this \emph{light sterile neutrino} (\lsn).
The squared-mass difference $\Delta m^2_{SBL}$ implies
that the sterile neutrino has a mass 
$m_s\simeq m_4\simeq \sqrt{\Delta m^2_{SBL}}\simeq1\eV$.
The opposite choice, that is 
an inverted sterile hierarchy with $m_4\ll m_1, m_2, m_3$, is forbidden by
the cosmological bounds on the neutrino masses \cite{Ade:2015xua}
and 
by the experimental bounds on
neutrinoless double-$\beta$ decay, 
assuming that massive neutrinos are Majorana particles
(see Ref.~\cite{Bilenky:2014uka}).

\section{Neutrino Parameterization}
To describe the contribution of the \lsn\ in Cosmology
we must separate the epoch when it was relativistic from the final
part of the Universe evolution, when it is non-relativistic.

The contribution of all the relativistic species to the energy
density of radiation $\rho_r$
in the early Universe can be written in terms
of the effective number of relativistic species \Neff,
defined through:
\begin{equation}
\label{eq:rho_rad}
  \rho_{r}
  =
  \left[1+\frac{7}{8}\left(\frac{4}{11}\right)^{4/3} \Neff\right]
  \rho_{\gamma}
  \,,
\end{equation}
being $\rho_{\gamma}$ the energy density of photons.
The contribution of the three active neutrinos is 
$\Neff^{\mathrm{sm}}=3.046$,
larger than 3 because of the non-instantaneous decoupling of the neutrinos
from the electron-photon fluid
and of the entropy transfer at the electron decoupling
\cite{Mangano:2005cc}.
If the \lsn\ is fully thermalized with the active neutrinos,
its contribution to \Neff\ should be 1 and $\Neff=4.046$.
Assuming that only photons and neutrinos
are relativistic in the early Universe,
the contribution of the \lsn\ can be written as \cite{Acero:2008rh}
\begin{equation}
 \DNeff
  =
  \Neff-
  \Neff^{\mathrm{sm}}
  =
  \left[\frac{7}{8}\frac{\pi^2}{15}{T_{\nu}}^4\right]^{-1}
  \frac{1}{\pi^2}
  \int dp \, p^3 f_{s}(p)
  \,,
\end{equation}
where $T_\nu$ is the active neutrino temperature,
$p$ is the neutrino momentum and
$f_s(p)$ is the \lsn\ momentum distribution function.

Since the neutrino temperature today is $T_\nu\propto 10^{-4}\eV$,
the \lsn\ is no more relativistic.
The energy density of the \lsn\ when it becomes non-relativistic
can be parameterized through the effective mass \meff{s},
defined as \cite{Acero:2008rh,Ade:2013zuv}:
\begin{equation}
\frac{\meff{s}}{94.1 \, \mathrm{eV}}
  =
  \frac{h^2}{\rho_c}
  \frac{m_{s}}{\pi^2}
  \int dp \, p^2 f_{s}(p)
  \,,
\end{equation}
where $h$ is the reduced Hubble parameter and
$\rho_c$ is the critical energy density.

Both \DNeff\ and \meff{s} depend on
the \lsn\ momentum distribution function $f_s(p)$.
If the \lsn\ is relativistic at decoupling,
$f_s(p)$ does not depend on its mass $m_s$.
Since sterile and active neutrinos are mixed, and since
the \lsn\ does not have electroweak interactions,
its decoupling cannot occur later than that of the active neutrinos,
that takes place at temperatures of about 1~MeV.
Depending on the production mechanism of the \lsn\, there are two
main possibilities for $f_s(p)$.
If the \lsn\ is produced through some thermal (TH) process,
its momentum distribution function can be written as a
Fermi-Dirac function at a temperature $T_s$, that is
$f^{TH}_{s}(p)=\left(e^{p/T_{s}}+1\right)^{-1}$.
In this case it is possible to show that
$\meff{s} = \DNeff^{3/4}\, m_{s}$.
If the production is non-thermal, instead,
there are several possible scenarios.
A popular one is the non-resonant production
or Dodelson-Widrow (DW) scenario \cite{Dodelson:1993je}.
In the DW scenario the \lsn\ momentum distribution function
can be written as 
$f^{DW}_{s}(p)=\beta\left(e^{p/T_\nu}+1\right)^{-1}$,
where $\beta$ is a normalization factor,
and one obtains
$\meff{s} = \DNeff\, m_{s}$.

\section{Neutrino Effects on the CMB spectrum}
\begin{figure}[t]
\centering
\includegraphics[height=8.5cm]{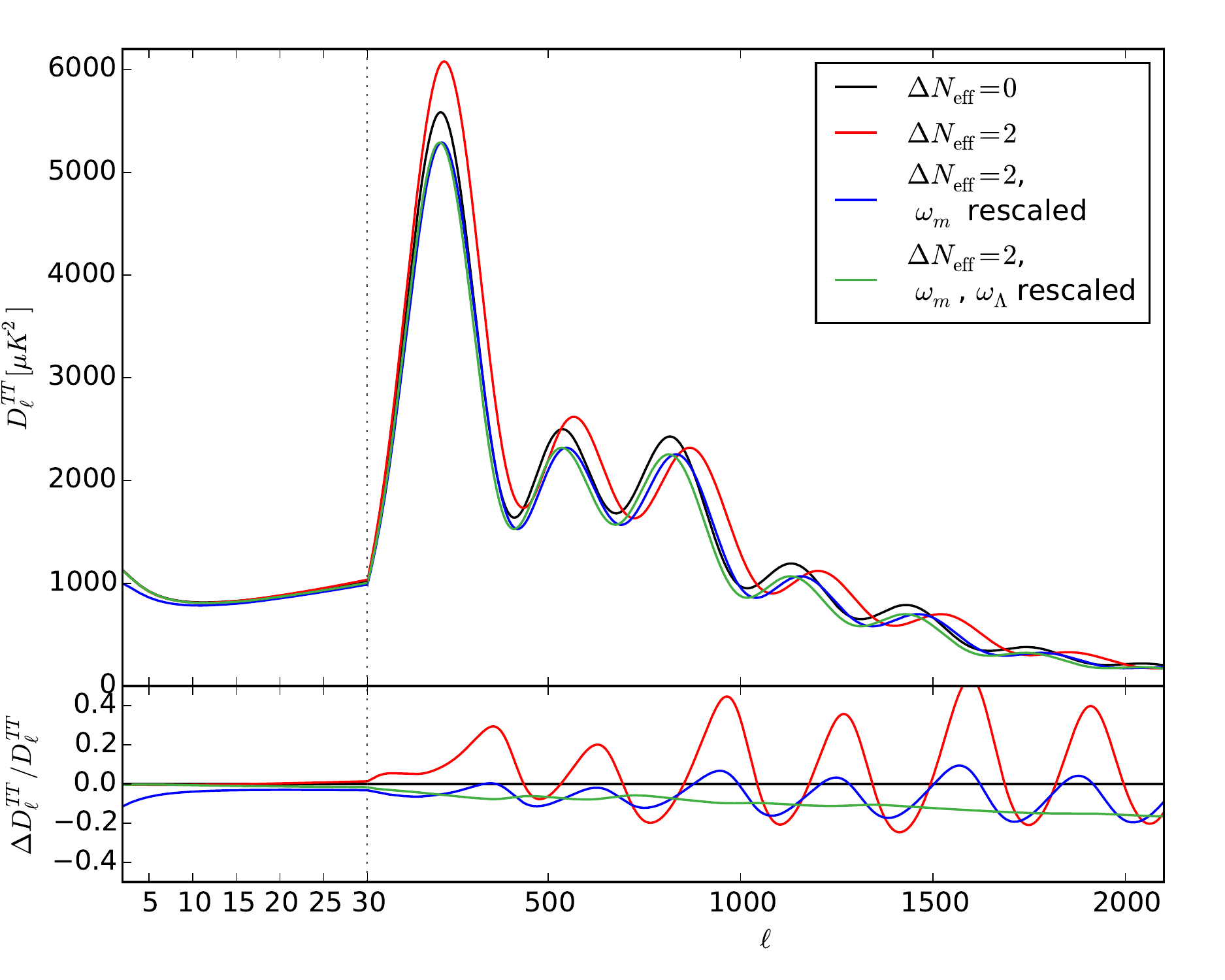}
\caption{Comparison of different CMB temperature spectra
predicted by the theoretical model.
We show the effects of altering \Neff\
alone or in combination with $\omega_m$ and $\omega_\Lambda$.}
\label{fig:neff_cl}
\end{figure}

The effective number of relativistic species 
controls the expansion rate $H$ in the early Universe
and the time of the matter-radiation equality.
This influences the CMB spectrum in several ways.
The redshift of matter-radiation equality is defined by
$1+z_{\mathrm{eq}}=\omega_r/\omega_m$,
where $\omega_i=\rho_i h^2/\rho_c$ is
the density parameter of the species $i$ and
the subscript $m$ is for the matter component.
If \Neff\ is increased, the equality is delayed and $H$ at CMB
decoupling is higher.
The consequences are \cite{Gariazzo:2015rra,Archidiacono:2013fha}
an increase of the first peak of the CMB,
due to the early Integrated Sachs-Wolfe (ISW) effect,
and a shift at higher
multipoles of the angular scale of the acoustic peaks.
These effects can be seen comparing the black ($\DNeff=0$)
and the red ($\DNeff=2$) spectra in Fig.~\ref{fig:neff_cl}.
If one wants to restore the matter-radiation equality to the former
redshift, it is necessary to rescale the matter density.
In order to maintain the redshift of the
matter-$\Lambda$ equality,
where $\Lambda$ indicates the cosmological constant, 
it is necessary to vary also the $\Lambda$ energy density.
If one varies together $\omega_m$ and \Neff\
(blue),
or $\omega_m$, $\omega_\Lambda$
and \Neff\ (green),
the effects of an higher \Neff\ can be partially compensated.
In the last case, the effect of an higher energy density
at all the times produces a suppression of the CMB spectrum
at high-$\ell$, also known as Silk damping effect.

As its mass is around 1~eV, the \lsn\ is still relativistic
at the time of matter-radiation equality.
As a consequence, its mass can influence only the late-time evolution.
In particular, the impact of $m_s$ on the CMB spectrum is visible
mainly through the slope of the CMB spectrum at low multipoles
and through the effect of neutrino free-streaming.
The slope of the CMB spectrum at low multipoles is
related to the late ISW effect, originated after the
Universe enters the $\Lambda$-dominated phase.
Neutrino free-streaming occurs since they have a large velocity and
they cannot cluster at small scales.
After the non-relativistic transition, however,
their velocity diminishes and they start to cluster,
as dark matter does.
The comoving free-streaming scale reaches a maximum at the time of
the non-relativistic transition, that occurs at the 
wavemode
$
k_{\mathrm{nr}}\simeq
0.0178
\,
\Omega_{m}^{1/2}
\left(T_{\nu}/T_{s}\right)^{1/2}
\left(m_{s}/1\eV\right)^{1/2} h \, \mathrm{Mpc}^{-1}
$.
All the wavemodes below $k_{\mathrm{nr}}$ are not affected by the
neutrino free-streaming, while all the wavemodes
$k\gtrsim k_{\mathrm{nr}}$ are suppressed.
The amount of the suppression depends on the neutrino masses,
but it is not particularly strong in the CMB spectrum.
The best way to measure the neutrino masses in Cosmology, in fact,
is to study 
the free-streaming effect in the power spectrum of matter perturbations.
Future experiments such as
DESI \cite{Levi:2013gra}
and Euclid \cite{Scaramella:2015rra}
will put strong constraints on the neutrino masses,
and possibly on the neutrino mass hierarchy
\cite{Font-Ribera:2013rwa,Audren:2012vy}.

\section{Current Constraints from Cosmology}
Several joint analyses of cosmological and SBL neutrino oscillation
data have been performed in the past.
Here we present the results obtained in Ref.~\cite{Gariazzo:2013gua},
where the joint analyses are based
on the CMB temperature spectrum from
the Planck 2013 release \cite{Ade:2013sjv},
on the low-$\ell$ polarization spectra from WMAP \cite{Bennett:2012zja}
and on the high-$\ell$ spectrum
calculated through the
ACT/SPT likelihood \cite{Dunkley:2013vu}.

\begin{figure}[t]
  \begin{minipage}{5.9cm}
     \centering
     \includegraphics[width=5.9cm]{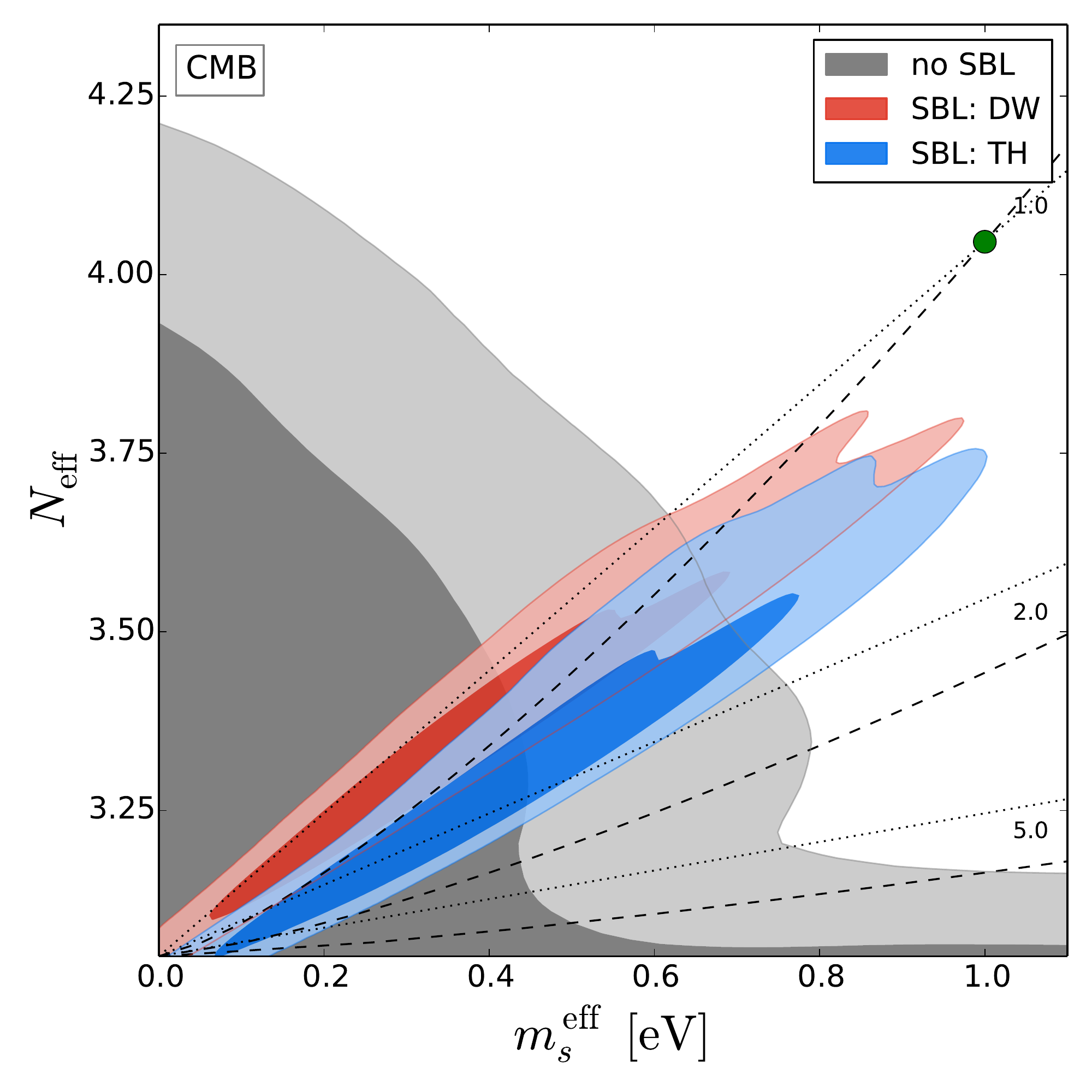}
  \end{minipage}
\hfill
  \begin{minipage}{5.9cm}
     \centering
     \includegraphics[width=5.9cm]{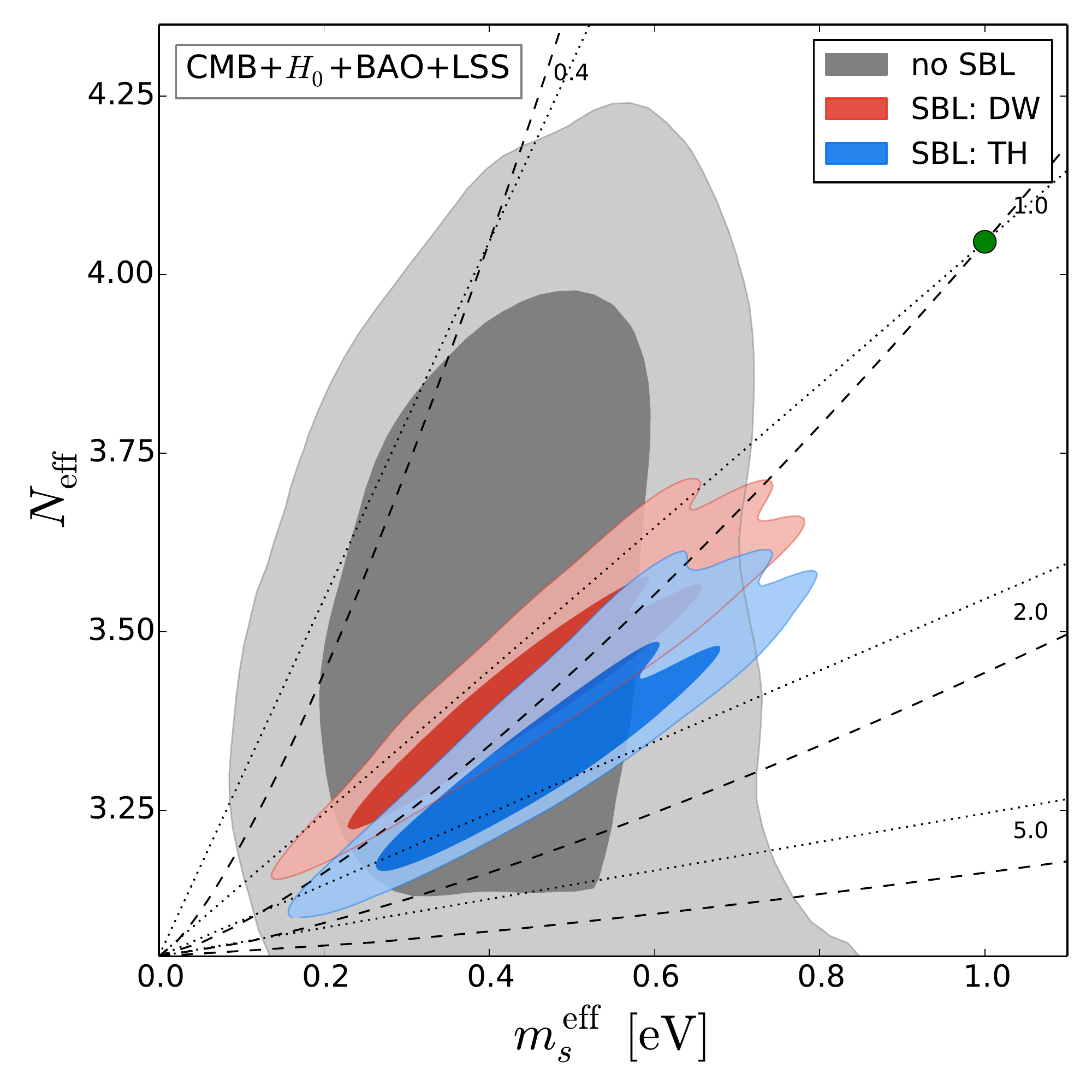}
  \end{minipage}
\caption{Constraints at 68\% and 95\% CL in the \meff{s}-\Neff\ plane
from the CMB data alone (left panel) and 
from a combination of CMB+$H_0$+BAO+LSS data (right panel).
The green point indicates a fully thermalized \lsn\ with $m_s=1\eV$.
Dotted (dashed) lines indicate constant masses,
for values written in the figures, for the DW (TH) model.
From \cite{Gariazzo:2015rra}.}
\label{fig:cosmo_sbl}
\end{figure}

To derive constraints on the \lsn\ properties
we consider an extension of the well-known \lcdm\ model,
expanded with the \lsn\ presented in the previous Sections.
The constraints are obtained from the cosmological datasets alone
or adding the results of 
the analysis of SBL neutrino oscillations data
as a prior for the \lsn\ mass.
In the left panel of Fig.~\ref{fig:cosmo_sbl}
we show the marginalized constraints in the \meff{s}-\Neff\ plane
obtained from the analysis of the CMB data mentioned above,
both without SBL prior (black)
or with a SBL prior for the mass of a DW (red) or a TH (blue) \lsn.
Dotted (dashed) lines indicate constant masses,
with values written next to each line, for the DW (TH) model.
As we can see, the SBL prior forces the physical mass to lie in the
range allowed by SBL experiments, with the consequence that not
all the values of \meff{s}-\Neff\ favoured by the CMB data are compatible
with the SBL data.
A fully thermalized \lsn\ is allowed
only if it has a very small mass, but for the SBL masses
an incomplete thermalization is required,
being $\Neff\lesssim3.75$ at 95\% CL.
Larger \lsn\ masses are allowed only if \Neff\ is close to 3.046.

The \lcdm\ model explains the majority of the
cosmological observations, but some tensions exist.
In particular, in the context of the \lcdm\ model
a tension between the
determinations of the matter fluctuations
and of the Hubble parameter at high- and low-redshift may appear,
possibly because of unaccounted systematics.
The value of $H_0$ determined by Planck \cite{Ade:2013zuv,Ade:2015xua}
is in tension with the local determinations, that give higher
values \cite{Riess:2011yx,Efstathiou:2013via}.
At the same time, the amount of matter fluctuations at small
scales, encoded in the parameter $\sigma_8$,
is smaller if obtained from local measurements of
weak lensing \cite{Heymans:2012gg}
or from cluster counts \cite{Ade:2013lmv}
than when derived from CMB constraints
\cite{Ade:2013zuv,Ade:2015xua}.

A \lsn\ has the potential to
reconcile local and cosmological observations.
This is the consequence of the correlation between \Neff\ and $H_0$,
and of the \lsn\ free-streaming, that reduces the matter fluctuations
at small scales.
In the right panel of Fig.~\ref{fig:cosmo_sbl} we show the bounds
for the \lsn\ as derived from an extended dataset,
that includes the same CMB data as above, plus
the measurements of the Baryon Acoustic Oscillations (BAO),
a prior on $H_0$
and
one on $\sigma_8$ from Large Scale Structures (LSS).
As we anticipated, the fit prefers positive \lsn\ masses
in order to reduce the predicted value of $\sigma_8$
through the neutrino free-streaming,
and $\Neff>3.046$ in order to have an higher $H_0$.
Even in this case, however, the full thermalization of the \lsn\
is strongly disfavoured.

\section{Open Problems and Conclusions}
Despite the fact that the \lsn\ can explain some of the tensions
between the current cosmological observations
at high- and low-redshift,
it seems that a full reconciliation is rather complicated.
The main reason is that we do not have a valid explanation
to the fact that the \lsn\ is not fully thermalized.
With the mixing parameters determined by SBL oscillations, in fact,
the \lsn\ should fully thermalize with the active neutrinos
\cite{Hannestad:2012ky}. 
Moreover, it seems that the \lsn\ cannot fully reconcile both
the $H_0$ and the $\sigma_8$ tensions,
since the direction of the degeneracies in the $\sigma_8$-$H_0$ plane
is not optimal \cite{Ade:2015xua}.

Unless a new mechanism that explains the low thermalization of
the \lsn\ is found (see e.g.\ Ref.~\cite{Gariazzo:2015rra}
for some possibilities),
the current status of the global analyses shows that
the SBL neutrino seems to be excluded by Cosmology.
Future investigations of SBL neutrino oscillations
and of Cosmology are required to grow a more complete knowledge
of the theories
and to explain of the present tensions between the 
different experimental probes.

\section*{Acknowledgments}
The work of S.\ G.\ is supported by
the Theoretical Astroparticle Physics
research Grant No.~2012CPPYP7
under the Program PRIN 2012 funded by the Ministero
dell'Istruzione, Universit\`a e della Ricerca (MIUR).


\end{document}